# High-temperature and high-pressure study on columbite structured ZnNb$_2$O$_6$

A. Tyagi[a], P. Botella[b],*, A. B. Garg[a,c],*, J. Sánchez-Martín[b], D. Díaz-Anichtchenko[b], R. Turnbull[b], S. Anzellini[b], C. Popescu[d], D. Errandonea[b]

[a]High Pressure and Synchrotron Radiation Physics Division, Bhabha Atomic Research Centre, Mumbai 400 085, India
[b]Departamento de Física Aplicada-ICMUV, MALTA Consolider Team, Universitat de Valencia, 46100 Valencia, Spain
[c]Homi Bhabha National Institute, Anushaktinagar, Mumbai 400094, India
[d]CELLS-ALBA Synchrotron Light Facility, Cerdanyola del Vallés, 08290 Barcelona, Spain

*Corresponding author: P. Botella
E-mail address: pablo.botella-vives@uv.es
* Alka B. Garg
Email Address: alkagarg@barc.gov.in

## Abstract

High-temperature and high-pressure experiments were conducted on columbite-type ZnNb$_2$O$_6$, reaching temperatures up to 873 K at ambient pressure and pressures up to 30 GPa at ambient temperature, respectively. Through systematic analysis employing synchrotron powder X-ray diffraction and Raman spectroscopy, we examined the crystal structure and phonon behavior. Within the specified temperature range, the orthorhombic phase of ZnNb$_2$O$_6$ (space group: *Pbcn*) demonstrated notable phase stability, with a thermal expansion coefficient similar to that of isomorphic compounds. Notably, a reversible phase transition was observed under compression at 10 GPa, with diffraction experiments indicating a shift to a monoclinic structure (space group *P2/a*), which remained stable up to 30 GPa. Changes in Raman modes, lattice parameters, and the unit-cell volume were monitored. A significant 2.5% discontinuity in the unit-cell volume at the phase transition pressure from orthorhombic to monoclinic suggests a first-order phase transition. The bulk moduli of the orthorhombic and monoclinic phases were estimated as 165(7) GPa and 230(9) GPa, respectively- We also found that both phases exhibit an anisotropic response to pressure. Furthermore, first-principles calculations support consistently with experimental observations.

## 1. Introduction

Columbite is a mineral composed primarily of iron, manganese, niobium, tantalum, and oxygen. It is commonly found in association with another mineral called tantalite, and together they form what is known as the columbite-tantalite series, also referred to as coltan. Columbite and tantalite are both important sources of niobium and tantalum, which are valuable metals used in various high-tech applications, particularly for the manufacturing of tantalum capacitors that are used for cell phones, personal computers, automotive electronics, and cameras. Columbite is a complex oxide mineral with the chemical formula $(Fe,Mn)(Nb,Ta)_2O_6$. This formula reflects its variable composition, where iron and manganese can substitute each other, as can niobium and tantalum.

Columbite-structured niobates with the general formula $ANb_2O_6$, where $A$ is a divalent cation, have attracted significant attention due to their intriguing crystal structures and properties [1]. The crystal structures and physical properties depend on the specific cation ($A$) present in the composition of the material. $ZnNb_2O_6$ is a compound that crystallizes in a columbite-type orthorhombic structure with the space group *Pbcn*. It has garnered significant attention from researchers due to its wide range of applications. These include its use in dielectric ceramics for microwave devices, solar cells, photoelectrochemical activity, and as additives to enhance the performance of energy storage compounds [2][3][4][5].

The properties of columbite-type niobates can be influenced by changing the divalent cation ($A$). The different electronic configurations of the cations induce a second-order Jahn-Teller (SOJT) distortion in the octahedra of *d*-transition metals, resulting in significant changes in the crystal structure [6][7]. In addition, external parameters such as pressure or temperature can modify the crystal structure resulting in changes in the physical, chemical and mechanical properties of the compound [8]. For example, Matar *et al.* [9], based on density-functional theory simulations, proposed that the mineral fordite ($SnNb_2O_6$, SG: *C2/c*) undergoes a pressure-induced metallization at about 23 GPa. Their calculations suggested two high-pressure (HP) structures: either the columbite-type ($CoNb_2O_6$, SG: *Pbcn*) or the trirutile-type structure ($FeTa_2O_6$, SG: $P4_2/mnm$). Furthermore, two naturally occurring columbite minerals, ferrocolumbite and manganocolumbite $(Fe,Mn)(Nb,Ta)_2O_6$, showed phase stability up to 7 GPa according to X-ray diffraction studies. In addition, it was proposed that the substitution of Mn for Fe can lead to a reduction in the bulk modulus of the material [7][10]. At 12 GPa, the end member columbite, $MnNb_2O_6$ undergoes a partially reversible phase transition, but no crystal structure analysis of the observed HP phase has been provided [11]. In addition, the manganotantalite member $Mn(Ta,Nb)_2O_6$ (SG: *Pbcn*) undergoes a phase transition around 9.5 GPa accompanied by an increase in the coordination of Mn from six to eight, and of Ta from six to seven. In this case, the low- and high-pressure phases are described by the same space group [12].

Compounds with the columbite-type structure, such as $ANb_2O_6$ (where $A$ can be Ca, Mg or Ni) and $NaNbO_3$, show similar behavior under pressure. Shiratori *et al.* studied the effect of pressure on microcrystalline $NaNbO_3$ and showed a phase transition at about 10 GPa, but did not provide a structure for the HP phase [13]. Later, Huang *et al.* identified the high-pressure phase of $MgNb_2O_6$ as a monoclinic structure at 10 GPa [14]. However, this claim is only supported by a "preliminary indexing" and the space group and unit-cell parameters were not provided. Zhou *et al.* suggested that the most likely high-pressure phase structure of $CaNb_2O_6$ at 9.5 GPa was monoclinic with a space group $P2_1/c$ [15]. More recently,

Karmakar *et al*. found a phase transition in $NiNb_2O_6$ at 9 GPa and proposed a monoclinic structure with a *P*2/*m* space group as the high-pressure phase [16]. According to Karmakar *et al*., the stability of these compounds seems to be determined by the $NbO_6$ octahedra. Their study showed that the distortion of the $NbO_6$ octahedra under pressure leads to the phase transition. According to this hypothesis, the behavior of these compounds under pressure is determined by the $NbO_6$ octahedra and not by the divalent metal cation [16].

Minerals of the columbite group occur as accessory phases in granitic pegmatite, which is an igneous rock and contains strategic rare metals. There are few investigations on these compounds under high-temperature conditions [17][18]. Tarantino *et al*. found no phase transition in natural columbites up to 900°C [19]. Mansurova *et al*. showed that the ferrocolumbite $FeNb_2O_6$ showed a phase transition from an orthorhombic to a tetragonal structure in a narrow range of temperature from 1630 K to 1693 K [20]. The pressure-temperature phase diagram of the manganotantalite $Mn(Nb, Ta)_2O_6$ was mapped by Liu *et al*. in the 300 to 800 K range, but no phase transition was reported, however, the Clapeyron slope between the low- and the high-pressure phase was $dP/dT = 0.0073$ GPa/K [12].

The above literature shows that the investigation of the high-pressure and high-temperature effects on the crystal structure of $ANb_2O_6$ compounds is an active field. However, despite numerous studies examining the structural changes of various compounds with the columbite-type structure under HP conditions, there is a significant gap in the research on their behavior under high-temperature conditions. Furthermore, the lack of a detailed structural characterization in many existing studies creates uncertainties about the exact nature of the structural transformations in high-pressure phases. This study will provide valuable insights into the phase transitions and structural stability of columbite-type niobates at high pressures and temperatures. The findings significantly contribute to the understanding of their properties and potential applications in various fields, including materials science and mineralogy.

2. **Experimental and Computational Methods**

Polycrystalline $ZnNb_2O_6$ was synthesized by a solid-state reaction. High-purity ZnO (99.9% purity) and $Nb_2O_5$ (99.9% purity) from Sigma-Aldrich were kept at 473 K for one hour to remove moisture or organic impurities. They were then weighed in a stoichiometric ratio and mixed thoroughly using a pestle and mortar. The well-mixed powder was pelletized using a hydraulic press and heated at 773 K for 24 h in a programmable muffle furnace from Therelek Furnaces Pvt. Ltd., India. This procedure was followed by a sintering at 1273 K for 36 h. The synthesized sample was characterized using a powder x-ray diffractometer from Proto Manufacturing Ltd. using copper $K_\alpha$ radiation ($\lambda$=1.5406 Å) and a silicon strip detector, confirming the formation of a single-phase of $ZnNb_2O_6$.

Raman spectra were acquired using a 750 mm focal length spectrometer coupled with a high-resolution charged coupled device (CCD) detector. A laser with a 532 nm wavelength was used for exciting the Raman signal. The entrance slit to the spectrometer was fixed at 50 µm. The laser power and exposure time were optimized to avoid sample degradation while maximizing the signal-to-noise ratio. Raman spectra under temperature variation were collected using a Linkam temperature stage, which provides controlled and stable temperature from 77 K to 873 K. For HP Raman measurements, we used a diamond-anvil cell (DAC) equipped with diamonds with culets of 300 µm in diameter. A stainless-steel gasket was indented to a thickness of 50 μm and a hole of 100 μm diameter was drilled in the center. Methanol-ethanol in a 4:1 ratio was used as a pressure-transmitting medium (PTM) and the ruby fluorescence method was used for pressure determination [21][22].

The effects of temperature on the crystal structure in the 300–800 K range were studied using a Philips X'Pert Pro Alpha1 diffractometer with a Bragg–Brentano geometry, working in continuous scanning mode, and using Cu $K_\alpha$ radiations and a HTK 1200 N (Anton Paar) temperature stage. HP powder angle-dispersive synchrotron XRD measurements were carried out at room temperature at the MSPD-BL04 beamline at the ALBA synchrotron [23]. A monochromatic X-ray beam of wavelength 0.4642 Å and beam size of 20 × 20 μm$^2$ was used. XRD patterns were recorded on a Rayonix CCD with a sample-detector distance of 300 mm which was determined using a $LaB_6$ standard. A rocking of ±3° of the DAC was used to reduce preferred orientation effects. We used a DAC with diamonds culets of 500 μm. The sample was loaded in a 200 μm diameter chamber drilled in a stainless-steel gasket, pre-indented to a thickness of 50 μm. The PTM was the same as in HP Raman studies. Special care was taken during sample loading to prevent sample bridging between diamond anvils [24]. The pressure was determined from XRD from a copper grain [25]. The data collected with the CCD detector were integrated into conventional diffractograms using Dioptas [26].

The refinement of the integrated XRD patterns was performed using the GSASII [27]. The Vesta software was used for structural representation [28]. The determination of the main axes of compressibility of the monoclinic structure and its linear compressibility was performed using Pascal [29]. In the Rietveld refinement the background of was described by a Chebyshev function with 10 coefficients. In the first refinement, only the background and the histogram scale factor were adjusted. In the second step, the lattice parameters and peak shape parameters were fitted.

DFT with a plane-wave basis set and pseudopotentials was used for performing first-principles calculations at zero temperature using Quantum Espresso [30]. The generalized-gradient approximation (GGA) prescribed by Perdew Burke-Ernzerhof was used for the calculation of exchange and correlation energies for the columbite structure of $ZnNb_2O_6$ [31]. The projected-augmented wave (PAW) scheme was employed as pseudo potential in self-consistent calculations [32]. Pseudopotentials for zinc, niobium and oxygen atoms were taken from the PSlibrary [33]. In the GGA approach, the electronic wave functions are expanded in a plane wave basis set with an energy cut-off of 90 Ry and a charge-density cut-off of 360 Ry to achieve highly converged results. A Monkhorst-Pack grid of 8×8×8 was used for Brillouin zone integrations. Structures reported in this paper were optimized at each pressure, from 0 up to 20 GPa with increments of 1 GPa, through the calculation of the forces and the stress tensor. At each pressure, geometric relaxation of the structure was achieved using the Broyden-Fletcher-Goldfarb-Shanno (BFGS) minimization scheme [34], the structure was fully relaxed to equilibrium conditions until the force on each atom was less than $3×10^{-5}$ eV atom$^{-1}$ so that the atoms were in local energy minima. A variable cell relaxation was performed on relaxed structures with a maximum deviation of the stress tensor from a diagonal hydrostatic form smaller than 0.01 GPa. After the equilibrium condition was obtained at each pressure, density-functional perturbation theory (DFPT) involving the calculation of second-order derivatives of energy was used to construct the dynamical matrix [35]. Normal vibrational frequencies at the Brillouin zone centre are deduced after the diagonalization of the dynamical matrix. The Raman modes have been assigned according to the calculations described above.

## 3. Results and discussion

**Ambient Structure**

The Rietveld refined X-ray diffraction data on the as-synthesized sample is shown in Figure 1a. All the observed peaks can be well indexed by the columbite-type orthorhombic structure with space group *Pbcn* [7] [16]. The structure is composed of two distinct octahedral units $ZnO_6$ and $NbO_6$ where the divalent $Zn^{2+}$ and pentavalent $Nb^{5+}$ cations occupy two distinct sites, *4c* and *8d*, respectively (see¡**Error! No se encuentra el origen de la referencia.**. 1b). The $NbO_6$ octahedra are connected via edge-sharing, thereby forming zig-zag chains along the *c*-axis. These chains are connected along [100] through the $ZnO_6$ octahedral units by corner-sharing. The $ZnO_6$ octahedral units also form zig-zag chains along the *c*-axis and the cations inside the octahedra are in the alternate sequence of -Zn-Nb-Nb-Zn-Nb-Nb- along the *a*-axis [36]. The refined unit-cell parameters are: $a$ = 14.2055(3) Å, $b$ = 5.7263(2) Å, $c$ = 5.0397(2) Å and $V$ = 409.91(2) Å$^3$. The goodness-of-fit parameters are $R_{wp}$ = 2.76 %, $R_p$ = 1.91 %, and $\chi^2$ = 2.1. The lattice parameters agree with those previously reported and are similar to the parameters of isomorphic compounds [2][12][16][20][37].

**High-temperature structural and vibrational studies**

The X-ray diffraction data collected up to 800 K are shown in Figure 2. All the data were indexed with the orthorhombic structure, indicating the structural stability of the compound under temperature. The diffraction data were analyzed using the Rietveld method to extract the variation in structural parameters. Since the high-temperature data were collected using a laboratory-based X-ray source, the data quality is not good enough to accurately refine all the atomic positions. Thus, only unit-cell parameters were refined and the atomic positions were not refined and taken from Waburg and Müller-Buschbaum [38]. The goodness-of-fit parameters suggest that the high-temperature data are of similar quality to those obtained at ambient temperature. The evolution of lattice parameters and unit-cell volume under high-temperature are presented in Fig. *3*. The temperature dependence of the unit-cell parameters follows a linear behavior according to the coefficient of determination ($R^2$ = 0.99) in the measured range. Thermal expansion coefficients (inset of Figure 3) were calculated from the linear fit of the Berman equation, $\alpha_x = 1/x_0(\delta x/\delta T)_P$, where x = *a*, *b*, or *c*, as implemented in EosFit software [39]. The observed thermal expansion coefficients show a good match with the previously reported values for natural columbites [19]. These findings are in line with an earlier report by Tarantino *et al.*, where it has been shown that the different ionic radii of divalent cations do not have a significant effect on the thermal expansion of columbites [19]. In $ZnNb_2O_6$ the volumetric thermal expansion coefficient is $\alpha_V$ = 25.0(5) × 10$^{-6}$ K. In this case, the order of axial thermal expansion coefficients is: α(c) > α(a) > α(b), which has been observed previously for natural ferrocolumbites. The maximum expansion corresponds to the direction of the zig-zag chains along the *c*-axis, indicating anisotropic expansion in the material [19][36]. The anisotropic expansion is also evident in the distortion of the ZnO and NbO polyhedra, each exhibiting distinct bond expansion rates, as depicted in Fig. *4* and detailed in Table 1 (see Fig. *5* for atom and bond identification). Bond distances and angles were calculated using VESTA. The bond length expansion demonstrates a linear trend, as indicated by the $R^2$ values in Table 1. However, the angles connecting the different metals deviate from this linear behavior, likely due to the anisotropic distortion of the polyhedra.

We will comment now on the reason for the anisotropy expansion of $ZnNb_2O_6$. As already observed in $MnTa_2O_6$ [36] our study suggest that it is a consequence of a lattice reorganization taking place on heating which will cause shear deformations. The smallest thermal expansion of the b-axis is connected to the pseudo-layered characteristic of the crystal structure of columbite. The structure consists of layers of slightly distorted hexagonal closed oxygen octahedral chains that are perpendicular to the *a*-axis. The $NbO_6$ octahedral units divide the edges by zigzag chains along the *c*-axis. Therefore, the expansion of octahedral units as temperature increases is mainly a consequence of the expansion of the octahedral. The agreement of the behavior of $ZnNb_2O_6$ with that of other columbites, suggest this a general comportment of this family of compounds.

In the following paragraph, we discuss the evolution of the Raman-active phonons of the compound with temperature. The primitive cell of $ZnNb_2O_6$ contains four formula units, which give a total of 36 atoms. These give rise to 108 phonons at the Brillouin zone center [40] which can be distributed in an irreducible representation of the $D_{2h}$ point group as:

$$\Gamma = 13A_g + 14B_{1g} + 13B_{2g} + 14B_{3g} + 13A_u + 14B_{1u} + 13B_{2u} + 14B_{3u}$$

Out of these modes, 54 are Raman active modes

$$\Gamma_{Raman} = 13A_g + 14B_{1g} + 13B_{2g} + 14B_{3g}$$

As shown in Fig. **6**, at ambient pressure and temperature conditions 34 out of 54 expected Raman modes were observed in the wavenumber range from 60 $cm^{-1}$ to 1000 $cm^{-1}$. The most intense mode at 891 $cm^{-1}$, identified as $A_g$ mode is the symmetric stretching mode of $NbO_6$ octahedra. The modes lying in the region between 380 $cm^{-1}$ to 800 $cm^{-1}$ correspond to the internal vibration of $NbO_6$ octahedra. The oxygen atoms in the $NbO_6$ octahedra can have three positions; "chain" ($O_3$) position which denotes the oxygen atoms connecting two $NbO_6$ octahedra belonging to the same chain, "bridge" ($O_1$) positions which denotes the oxygen atoms connecting two successive $NbO_6$ octahedra and "terminal" ($O_2$) bonding to one Nb atom to two Zn atoms (see Fig. **5** for atom and bond identification) [41]. The Raman modes observed below 100 $cm^{-1}$ are due to lattice vibrations.

Fig. **7**a depicts the Raman data collected from 78 K to room temperature while

Fig. **7**b shows the data from ambient temperature to 873 K. The data collected at 78 K (the lowest temperature reached in the present study) shows 37 Raman modes as compared to 34 modes at ambient conditions. The extra three modes at 318, 429 and 496 $cm^{-1}$ are due to a reduction in FWHM of the closely lying modes resulting in their separation. On increasing the temperature from 78 K, all modes soften until 493 K due to the expansion of the lattice. Two extra Raman modes start appearing at 233 and 879 $cm^{-1}$ at 493 K, which remain up to the highest temperature reached in the present studies. The strongest peak at ~ 900 $cm^{-1}$ is associated with the Nb-O2 bonds and the strongest peak at ~ 530 $cm^{-1}$ is associated with the Nb-O1 bonds [15]. Under high-temperature, the Nb-O2 bonds are less affected by temperature (see Table 1) and the strongest peak can be followed up to the maximum temperature reached. On the other hand, the ~ 530 $cm^{-1}$ vibration mode is more sensitive to temperature and becomes broader at around 653 K, but still can be followed. Despite the appearance of new modes, which is indicative of a possible phase transition, no phase transition is observed in XRD measurements.

Fig. 8a illustrates the temperature dependence of the Raman modes up to 370 cm$^{-1}$ and

Fig. 8b displays modes from 370 cm$^{-1}$ to 900 cm$^{-1}$. The non-linear red shift in Raman modes with temperature from 78 K to 873 K indicates that the predominant contribution to the anharmonicity in ZnNb$_2$O$_6$ arises from the three-phonon decay process [42]. The non-linearity is more important in low-frequency lattice modes that in high-frequency internal octahedral modes, which is a consequence of the fact that octahedral are nearly rigid units.

### High-pressure structural and vibrational studies

The evolution of the X-ray diffraction patterns with pressure is shown in Fig. 9. The maximum pressure achieved during the experiment was 29.1 GPa. All the observed diffraction peaks are well indexed by the orthorhombic structure with space group *Pbcn* up to 10 GPa. The lattice parameters obtained with the sample already loaded inside the DAC (P = 0.1 GPa) are, $a$ = 14.1890(5) Å, $b$ = 5.7216(2) Å, $c$ = 5.0315(2) Å, $V$ = 408.49(2) Å$^3$ and the goodness-of-fit parameters are $R_{wp}$ = 15.44 %, $R_p$ = 8.54 %, and $\chi^2$ = 3.3 (see Fig. 10). The obtained unit-cell parameters agree well with those reported in the literature [2][12][16][20][37]. As the sample pressure is increased beyond 10 GPa (

Fig. 9), several changes can be observed in the XRD patterns. One of the most striking observations is the emergence of a new reflection adjacent to the initially observed most intense peak. Subsequently, with a further increase in the pressure, this newly observed reflection surpasses the original peak in intensity. A few extra reflections also appeared at low angles between 6º to 8º and at high angles between 10º to 12º, whilst reflections from the low-pressure phase could still be observed. Thus, a phase coexistence between the low- and high-pressure phases takes place between 10 and 13.2 GPa. A similar phase coexistence was observed for MnNb$_2$O$_6$, MgNb$_2$O$_6$, CaNb$_2$O$_6$, and NiNb$_2$O$_6$ [11][14][15][16]. On further increasing the pressure, only the high-pressure phase is observed until the highest pressure is reached in these measurements. The initial orthorhombic structure was completely recovered once the pressure was released.

After conducting an indexing procedure using GSAS [27], we found that the XRD pattern of the high-pressure phase could be indexed by a monoclinic structure described by space group *P*2/a (see Fig. 10). Notice that *P*2/*a* is equivalent to *P*2/*c* (obtained by the transformation (abc) to (c-ba)) which is a translationgleiche subgroup of space grou`*Pbcn*. After a Le Bail fit, the lattice parameters *a*, *b*, *c*, *β* and unit-cell volume of the monoclinic structure at 16.5 GPa were estimated as: $a$ = 9.037(2) Å, $b$ = 6.475(2) Å, $c$ = 7.183(7) Å, 118.43(2)º and $V$ = 369.7(2) Å$^3$, respectively. The corresponding goodness-of-fit parameters are $R_{wp}$ = 27.5 %, $R_p$ = 15.5 %, and $\chi^2$ = 3.2. Similar unit-cell parameters were obtained for the high-pressure phase of NiNb$_2$O$_6$ but the structure is described as monoclinic with space group *P*2/*m* [16]. Notice that the study of NiNb$_2$O$_6$ and our study are the only two studies where the symmetry and space group are proposed for the HP phase of a columbite. The two studies agree in a reduction of symmetry at the phase transition, which is sluggish and involves a collapse of the unit-cell volume, which suggests a reconstructive mechanism for the phase transition. HP single-crystal XRD [43] are necessary to fully solve the crystal structure of the HP phase of columbite. We hope our study will trigger single-crystal XRD studies.

The evolution of the unit-cell parameters and unit-cell volume obtained from XRD measurements under pressure are shown in Fig. 11. The linear behavior of lattice parameters agrees with our theoretical calculations for the orthorhombic phase and matches with the earlier results obtained for natural columbites studied by Tarantino *et al.* up to 7 GPa [7]. For the orthorhombic phase, the linear compressibility of various axes is estimated to be $\kappa_a = 1.28(5) \times 10^{-3}\ GPa^{-1}$, $\kappa_b = 2.39(3) \times 10^{-3}\ GPa^{-1}$ and $\kappa_c = 1.54(3) \times 10^{-3}\ GPa^{-1}$. The obtained values are comparable to that of natural columbites ($\kappa_a = 1.6 \times 10^{-3}\ GPa^{-1}$, $2.8 \times 10^{-3}\ GPa^{-1}$ and $\kappa_c = 1.7 \times 10^{-3}\ GPa^{-1}$) [7]. However, significant differences can be appreciated when compared to $NiNb_2O_6$ which shows higher values for the compressibility of the lattice parameters ($\kappa_a = 2.7 \times 10^{-3}\ GPa^{-1}$, $2.6 \times 10^{-3}\ GPa^{-1}$ and $\kappa_c = 2.4 \times 10^{-3}\ GPa^{-1}$) [16]. The $b$-axis is the most compressible and it is perpendicular to the zig-zag chain. This suggest that under pressure, the zig-zag chains might become aligned along the $c$-axis. Interestingly, the $b$-axis is the axis with the lowest thermal expansion, which means that thermal expansion and compression are governed by different mechanisms, which contrast with the behavior reported for more simple oxides [44].

Fig. 12 illustrates the evolution of ZnO and NbO polyhedral bonds and angles under varying pressure conditions. Up to 4.5 GPa, there is a noticeable trend of bond distance reduction. Beyond this pressure threshold, a significant distortion of the polyhedra becomes apparent, with distortion intensifying notably after reaching 7.5 GPa. This observation aligns with the subsequent discussion on the equation of state fit, where the deviation of the last three pressure points from the 3rd order Birch-Murnaghan equation is addressed.

For the HP monoclinic phase, as the compressibility tensor is not diagonal, the main axis of compression and its compressibilities have been determined using Pascal [29]. The main compressibility axes are [010], [102], and [001], and their respectable compressibilities are $\kappa_1 = 0.92(9) \times 10^{-3}\ GPa^{-1}$, $\kappa_2 = 1.01(8) \times 10^{-3}\ GPa^{-1}$, and $\kappa_3 = 0.9(3) \times 10^{-3}\ GPa^{-1}$. Thus, the least compressible axes are the $a$- and $c$-axis with similar compressibility values. The most compressible axis is the $b$-axis. However, the difference in compressibilities is of the order of 10%, indicating that the response of the crystal structure to external pressure is only slightly anisotropic. This observation is consistent with the reconstructive mechanism proposed for the phase transition. It is also observed that the $\beta$-angle under pressure follows a linear behavior, following pressure dependence (d$\beta$/d$P$ = 0.016(2) °/GPa) (see inset, Fig. 11 (right)).

The evolution of the unit-cell volume for the orthorhombic phase can be well fitted with a 2$^{nd}$ or 3$^{rd}$-order Birch-Mumaghan (BM2 or BM3) equation of state (EoS) (green and red lines in Fig. 11, right). For an accurate determination of the bulk modulus, the BM3 is considered the most suitable, as illustrated in the Eulerian strain-normalized pressure plot shown in Figure S1 in the supplementary material. While the fourth-order equation (BM4) appears to pass through more data points, it exhibits a negative pressure derivative, which implies that the material becomes more compressible as density increases, being not possible. In the plot of observed pressure versus calculated pressure (Pobs-calc), the BM3 EoS demonstrates a closer alignment with the reference line for pressures below 7 GPa, with only minor deviations observed at the last three pressure points. Table 2 (and Figure S2 in the supplementary material, confidence ellipses) presents data from reference [16] and the current study for comparing the unit cell volume, bulk modulus, and its derivative between experimental and calculated values, as well as with other columbite-type structures. Bulk moduli obtained using BM2 or BM3 (159.5(19) and 165(7), respectively) show similar values which are slightly higher than the natural columbite studied by Pistorino *et al* [10] and the $MnNb_2O_6$ studied by F. Huang *et al* [11]. On the other hand, the bulk modulus obtained is smaller than that of $NiNb_2O_6$ [16]. The substitution of the A cation within the columbite structure of $ANb_2O_6$, along with the degree of structural order, significantly influences the compressibility behavior of columbites [10]. In fact, the bulk modulus of columbite-type oxide can be expressed by means in terms of cation oxide polyhedral compressibilities as happens in spinel-type oxides [45] and other ternary oxides like zircon or scheelite [46]. In particular, if the model proposed in Refs. [45][46] values of the bulk modulus of 155 and 165 GPa are obtained supporting the results obtained from our experiments. To conclude this part of the discussion we will compare experiments and DFT calculations. The smaller values of bulk modulus obtained through DFT calculations (see Table 2) are in alignment with the observed trend of overestimation in the unit cell volume. This is a typical limitation of DFT related to the approximations used to describe the exchange-correlation [47].

An important thermodynamic magnitude is the product $α_VB_0$ [48] which to a first approximation remains constant with changing temperature along isobars [49] when the temperature is higher than the Debye temperature. For $ZnNb_2O_6$, the value of $α_VB_0$ = 0.0041(5) GPa/K, which is comparable to the value reported for many oxides including $Al_2O_3$, $MgO$, $MgSiO_3$, $Mg_2SiO_4$, olivine, pyrope, and garnet [50]. Since in the case of columbites the Debye temperature is below 300 K [51]. The value of $α_VB_0$ here reported can be used to determine the pressure dependence of the volume following different isotherms [52].

At the phase coexistence transition (green shadow area in Fig. 11), the unit volume exhibits a discontinuous drop of 2.5 %, which is indicative of a first-order phase transition. It can be observed that the two phases affect each other, as in the coexistence region, the volume of the orthorhombic phase shows larger values and the monoclinic phase shows a plateau before compressing. The bulk modulus of the high-pressure phase was determined to be 230(9) GPa using BM2. The value is consistent with that of the high-pressure phase of $NiNb_2O_6$ ($B_0$ = 244(6) GPa) [16]. As it has been discussed in the introduction section, reports on the high-pressure structural and vibrational studies on this family of compounds are sparse, with literature limited to $CaNb_2O_6$ [15] and $NiNb_2O_6$ [16]. In these studied compounds only the monoclinic high-pressure phase has been reported, however the structures have not been fully solved. Therefore, a systematic theoretical study of the high-pressure phase of $ANb_2O_6$ polymorphs is encouraged. Our study, also supports the symmetry decrease of the crystal structure of the HP phase. HP single-crystal XRD experiments could make a significant contribution to solving the structures of the HP phases.

The high-pressure phase transition was also detected by Raman spectroscopy measurements. In Fig. 13, the pressure-evolution of the Raman spectra is shown up to 31.6 GPa. Under almost ambient conditions inside the DAC (0.1 GPa), 32 Raman modes out of 54 expected modes have been observed for the orthorhombic structure (see Fig. 14 and Table 3). All of the calculated Raman, IR, silent and acoustic modes, as well as the respective mode symmetries, are reported in the Supplementary Material (Fig. S1). Several changes are observed with the increase in pressure. Following the application of pressures at 4.5 and 8.5 GPa, notable changes in the modes are observed. Specifically, the mode at 200 cm$^{-1}$ exhibits a merging phenomenon. Additionally, the broad features at 240 cm$^{-1}$ and the broad features at 400 cm$^{-1}$ undergo splitting into multiple modes, while certain other modes exhibit a merging behavior. This may be due to the anisotropic response of Raman modes to pressure as can be seen from the slopes in Fig. 15 and Table 3. It can be observed from Table 3 that all the modes harden under compression assuming a linear behavior for the pressure coefficients ($∂ω/∂P$). The Gruneisen parameter is also estimated for the experimental and theoretical data as $γ = (B_0/ω_0)(∂ω/∂P)$ using $B_0$ = 159.5(17) GPa. The most intense peak at 900 cm$^{-1}$, which is related to the Nb-$O_2$ (terminal oxygen, $O_t$), becomes weaker in intensity. After 10.1

GPa, new features can be seen in the Raman spectrum. A broad feature appears around 700 cm$^{-1}$ and several broad overlapped features due to the phase coexistence between the low- and high-pressure phases can be seen up to 15.9 GPa. Beyond this pressure, the Raman spectra correspond only to the high-pressure phase. Raman spectroscopy is more sensitive to structural changes induced by pressure, but the transition phase is determined to take place between 10 and 12 GPa which agrees with the XRD analysis albeit showing a larger range of phase coexistence. The Raman spectrum of the high-pressure phase can be fitted using 38 modes (see Fig. 14). It is essential to note that due to a lack of information on the complete atomic arrangement, the determination of the expected number of Raman modes for this phase remains a challenge for future work.

## 4. Conclusions

In summary, our comprehensive study of the structural and vibrational properties of $ZnNb_2O_6$ under high pressure and temperature has provided valuable insights into the behavior of the studied material under different thermodynamic conditions. The confirmation of a stable orthorhombic structure at ambient conditions, coupled with the observation of structural stability up to 800 K, underscores the robustness of the crystal structure in the studied compound. In particular, our analysis shows that the variance in ionic radii of divalent cations does not have a significant effect on thermal expansion. Under high pressure, the emergence of a monoclinic phase and the occurrence of phase coexistence reveal the propensity of the material to undergo a reconstructive structural transformation under extreme conditions. The identification of a first-order phase transition and the determination of bulk modulus values serve as key contributions that enrich our understanding of the high-pressure behavior of $ZnNb_2O_6$. By comparing our results with existing literature, we provide critical insights into the broader context of the $ANb_2O_6$ family of compounds, filling important gaps in current understanding. We expect that our study will open future research avenues could delve deeper into elucidating the crystal structure of the high-pressure phase and exploring its vibrational dynamics, offering promising prospects for applications in materials science and solid-state physics.

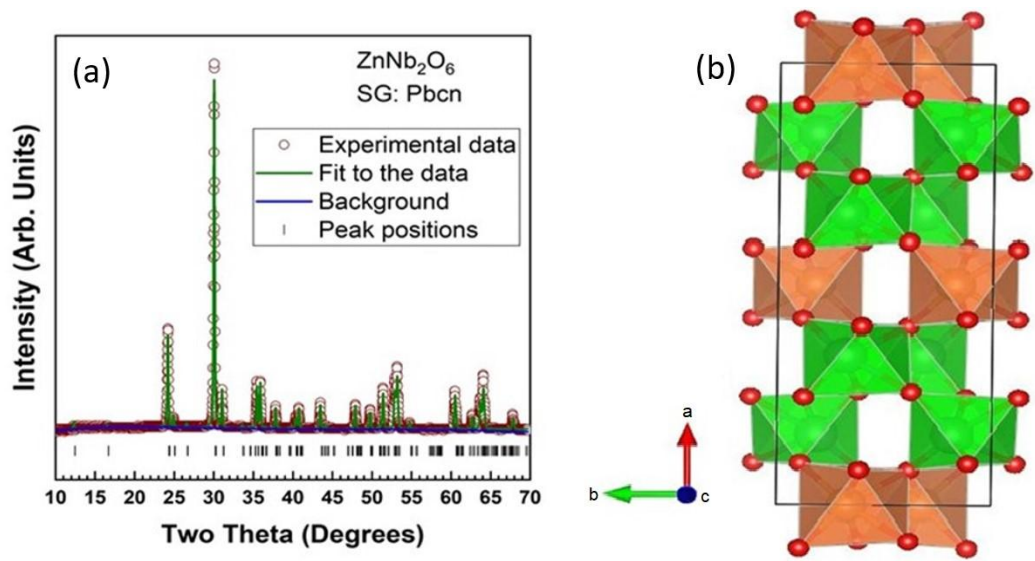

**Fig. 1.** (a) Rietveld refinement of the as-synthesized $ZnNb_2O_6$ XRD pattern acquired at ambient pressure and temperature. The results showed a single-phase formation of the compound. The fitted background plot is also included. The vertical tick marks represent the allowed reflections of the columbite orthorhombic structure with the *Pbcn* space group. (b) Representation of the columbite structure with oxygen octahedra around zinc ($ZnO_6$) and niobium ($NbO_6$) atoms in green and orange, respectively.

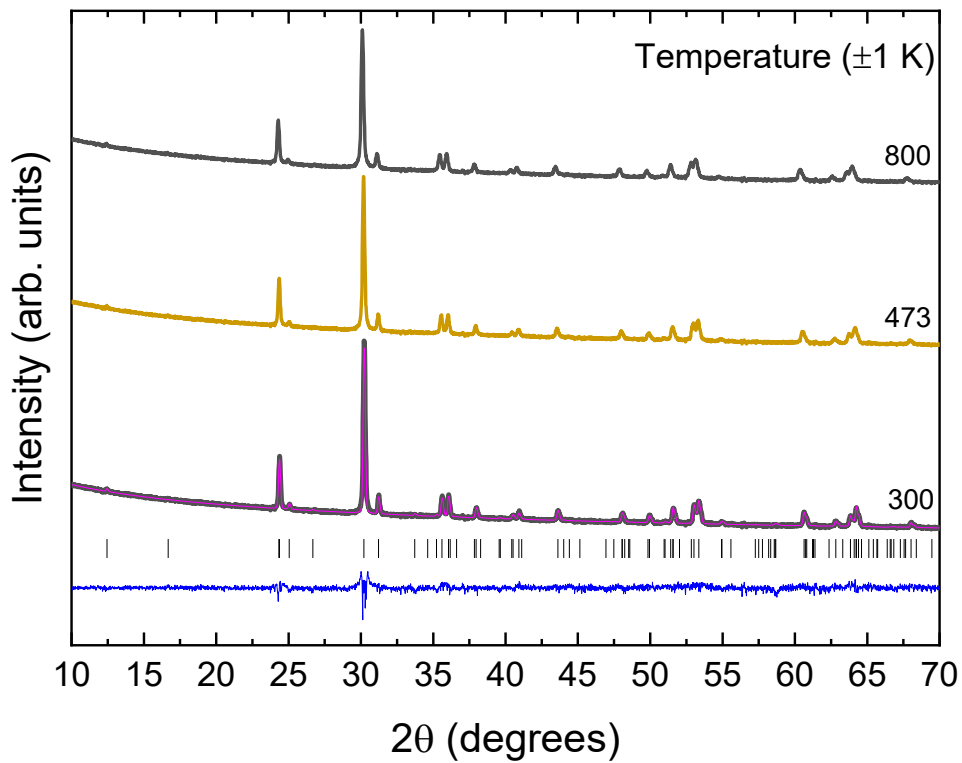

**Fig. 2.** Selected $ZnNb_2O_6$ X-ray diffraction data acquired during heating (P = 1 bar). The magenta line (color online) shows the Rietveld refinement for the sample measured at 300 K (Comparable refinements have been attained for the remaining patterns). Residuals are shown below for the adjusted pattern. Ticks correspond to the Bragg peak positions. Numbers on the right-hand side of the figure denote the temperature in Kelvin.

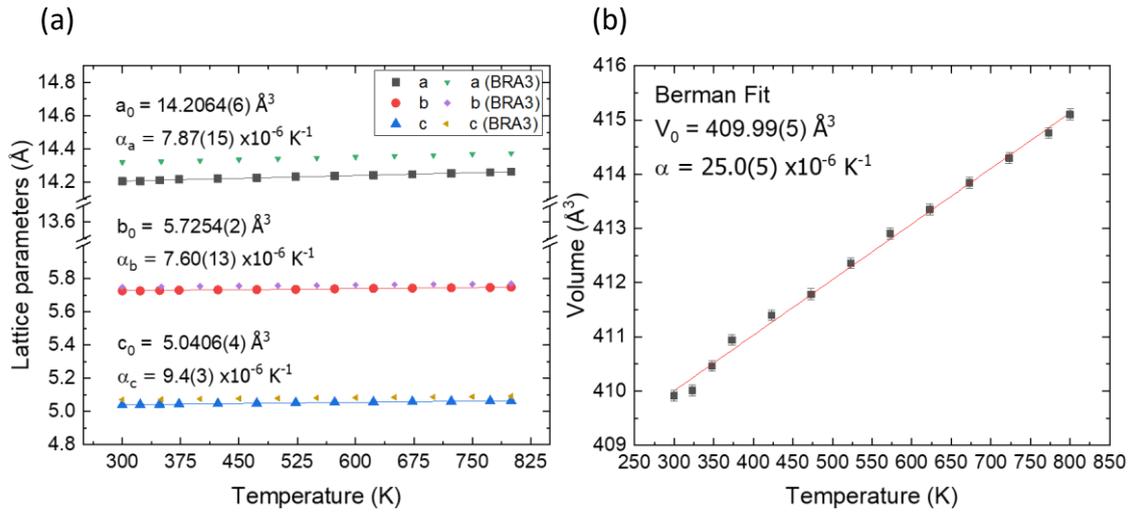

**Fig. 3.** (a) $ZnNb_2O_6$ lattice parameters evolution under high-temperature conditions (P = 1 bar) (error bars are included but they are smaller than the symbols). (b) $ZnNb_2O_6$ unit-cell volume expansion under high-temperature conditions. The parameters shown inside the plots are the reference values for the respective parameters at 300 K. (BRA3) data from ref. [19].

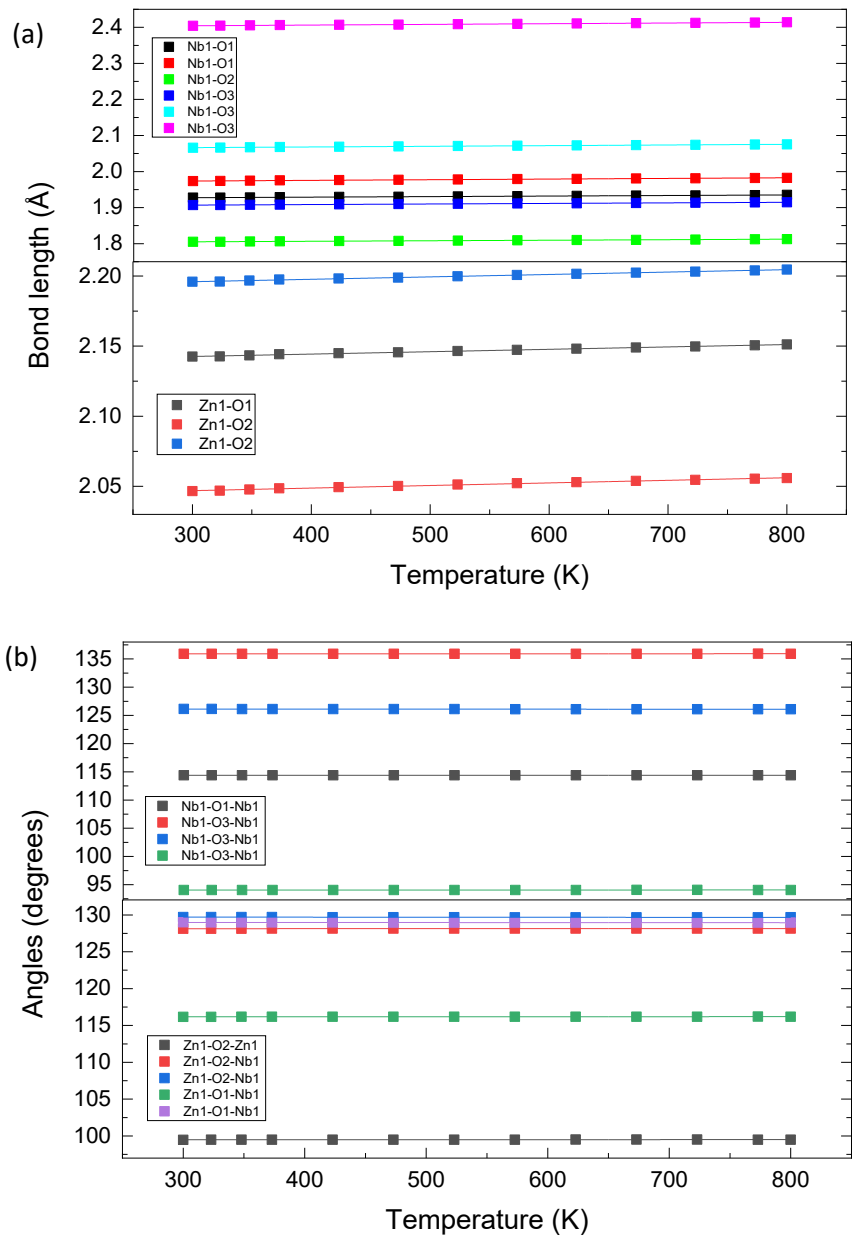

**Fig. 4.** (a) Bond length and (b) angles evolution for the two ZnO and NbO polyhedra under temperature at P = 1 bar.

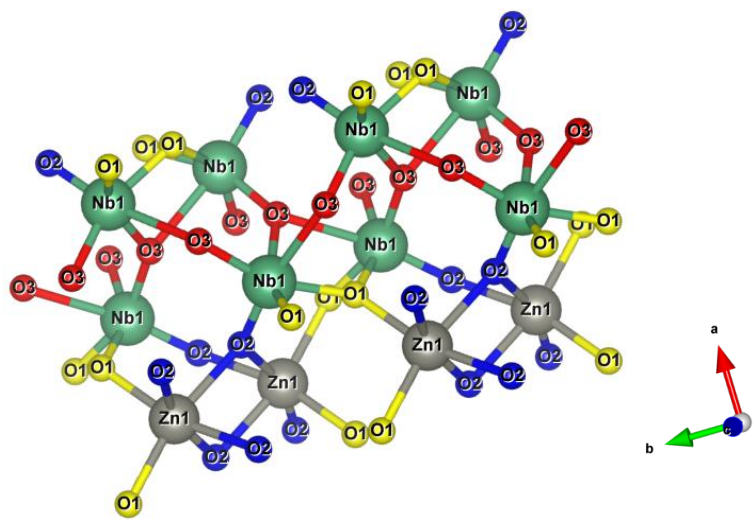

**Fig. 5.** Schematic representation of the columbite crystal structure showing the different atoms and bonds.

**Table 1:** Linear fit parameters from Fig. 4.

|  | Initial length (Å) | Slope (Å/K) $10^{-5}$ | $R^2$ |
|---|---|---|---|
| **Zn1-O1** | 2.143 | 1.71(3) | 0.997 |
| **Zn1-O2** | 2.047 | 1.84(5) | 0.993 |
| **Zn1-O2** | 2.196 | 1.72(3) | 0.997 |
| **Nb1-O1** | 1.927 | 1.51(2) | 0.997 |
| **Nb1-O1** | 1.973 | 1.80(5) | 0.992 |
| **Nb1-O2** | 1.805 | 1.45(2) | 0.997 |
| **Nb1-O3** | 1.907 | 1.55(3) | 0.997 |
| **Nb1-O3** | 2.066 | 1.83(4) | 0.993 |
| **Nb1-O3** | 2.404 | 1.89(3) | 0.997 |

|  | Initial angle (°) | Slope (°/K) $10^{-5}$ | $R^2$ |
|---|---|---|---|
| **Zn1-O2-Zn1** | 99.487 | 4.1(6) | 0.800 |
| **Zn1-O2-Nb1** | 128.132 | 3.9(6) | 0.774 |
| **Zn1-O2-Nb1** | 129.710 | -8(1) | 0.821 |
| **Nb1-O1-Nb1** | 114.373 | 4.5(7) | 0.783 |
| **Nb1-O3-Nb1** | 135.904 | -8(1) | 0.821 |
| **Nb1-O3-Nb1** | 126.128 | 2.9(4) | 0.787 |
| **Nb1-O3-Nb1** | 94.041 | 3.6(6) | 0.768 |
| **Zn1-O1-Nb1** | 116.171 | -8(1) | 0.822 |
| **Zn1-O1-Nb1** | 128.988 | 4.4(6) | 0.799 |

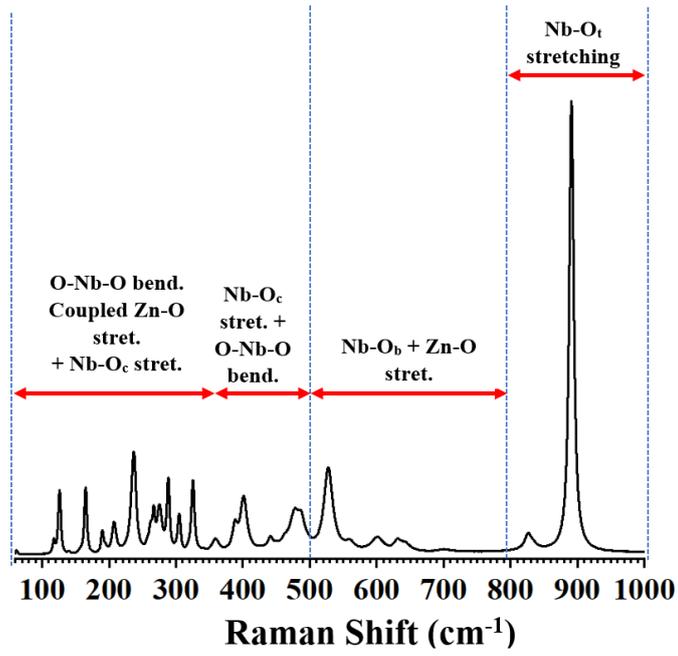

**Fig. 6.** Ambient conditions (T = 300 K, P = 1 bar) Raman spectrum of ZnNb$_2$O$_6$ showing three distinct vibrational regions.

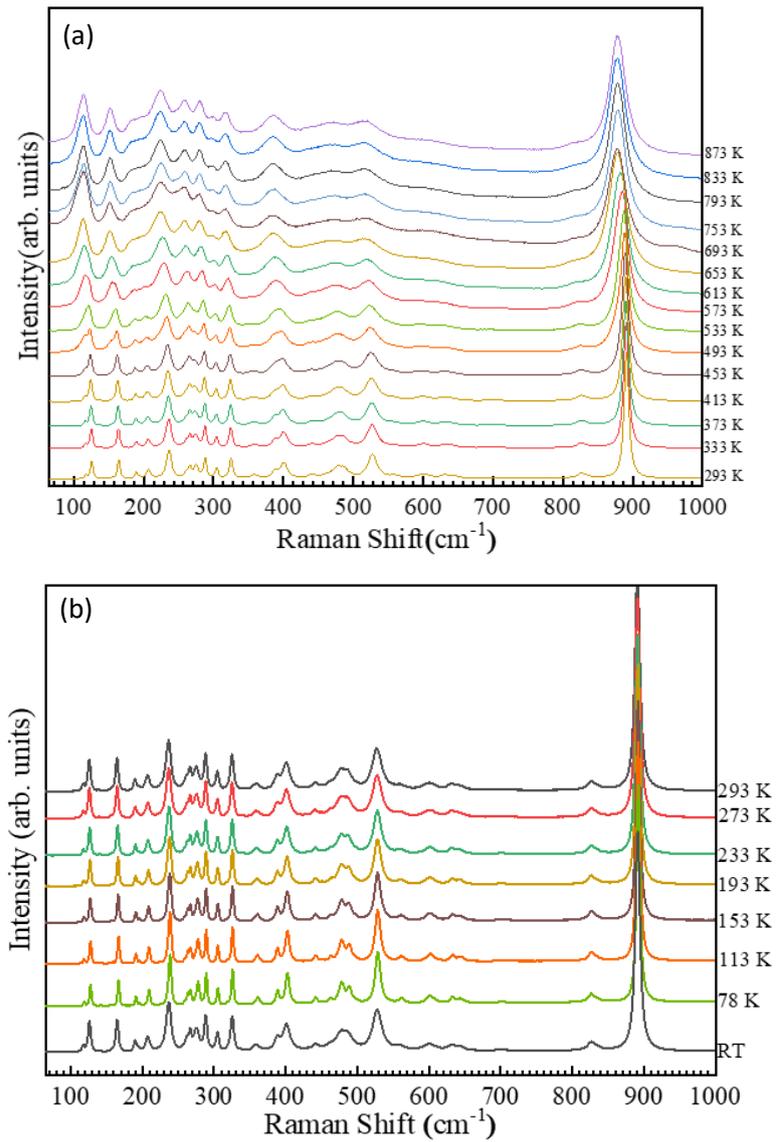

**Fig. 7.** (a) Raman spectra of $ZnNb_2O_6$ from 78 K to room temperature (RT). (b) From room temperature to 873 K. Measurements at P = 1 bar.

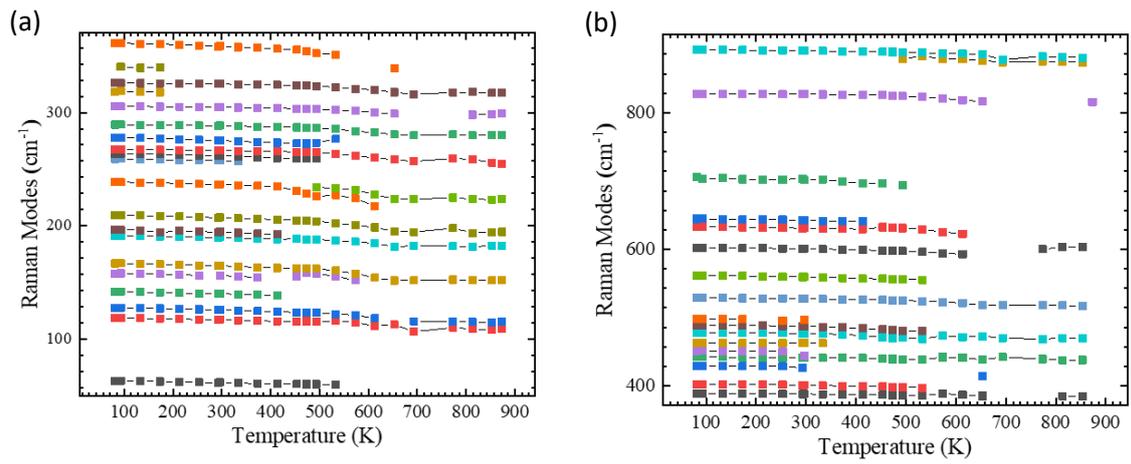

**Fig. 8**. Temperature variation of the $ZnNb_2O_6$ Raman modes: (a) from 60 to 370 cm$^{-1}$, (b) from 370 to 900 cm$^{-1}$. Measurements at P = 1 bar.

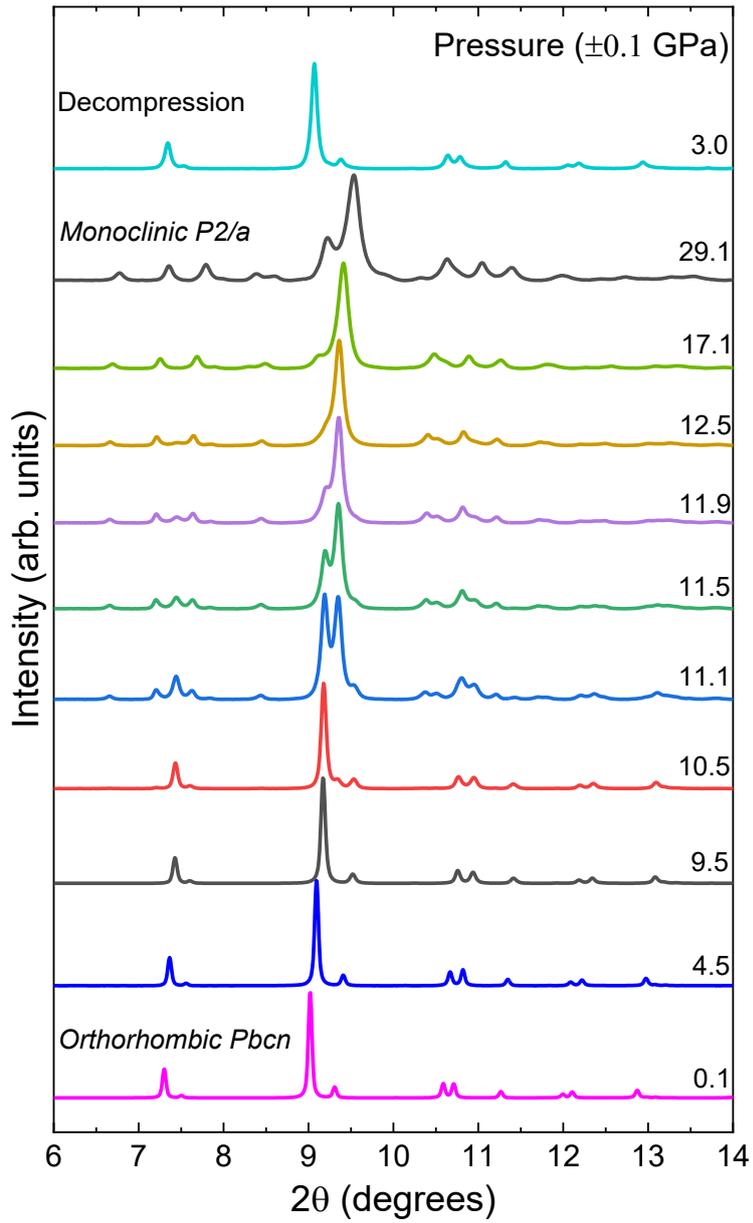

**Fig. 9**. Selected HP XRD patterns of $ZnNb_2O_6$. The recovered XRD pattern after releasing pressure is shown on top. Measurements at T = 300 K.

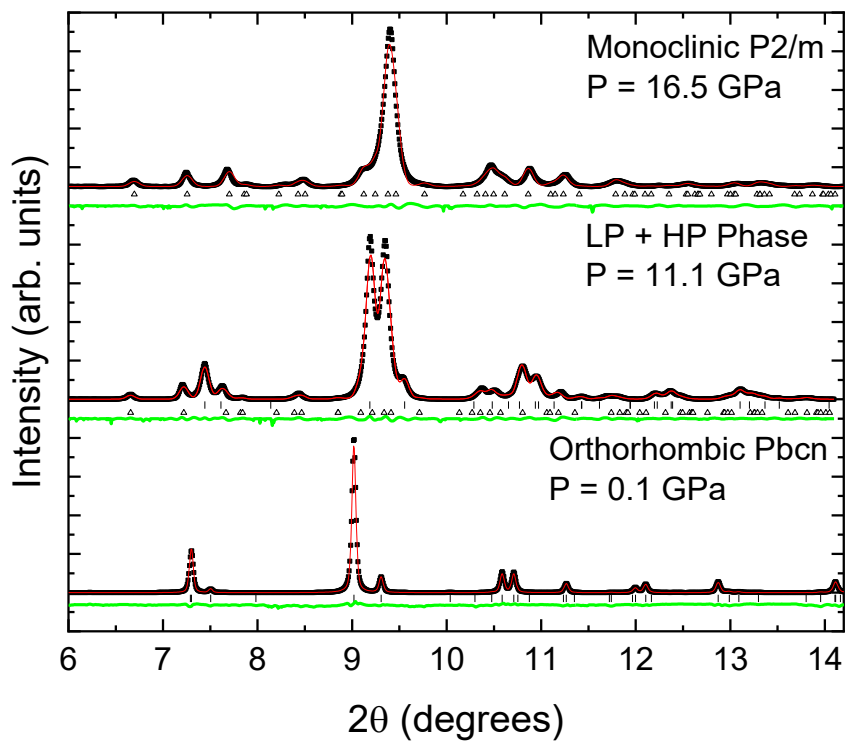

**Fig. 10.** Red line (color online) shows the Rietveld (LP)/Le Bail (HP) refinement for each $ZnNb_2O_6$ phase. Residuals are shown below for the adjusted patterns. Ticks correspond to the Bragg peak positions. Measurements at T = 300 K.

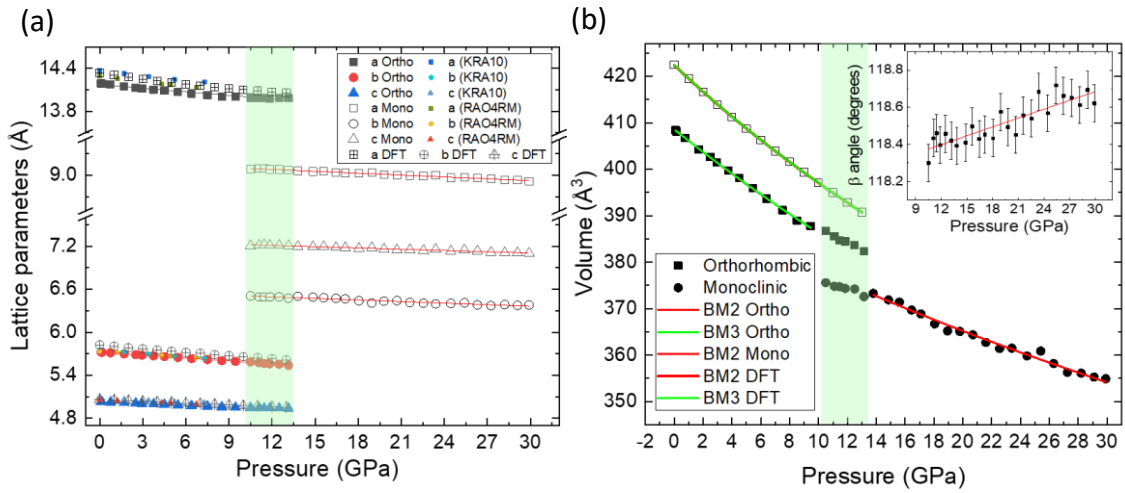

**Fig. 11.** (Color online) (a) Evolution of the ZnNb$_2$O$_6$ lattice parameters under pressure. The error bars are smaller than the symbols (σ(P) ~0.1 GPa, σ(parameters) ~0.001 Å). Lines represent linear fits. (b) Unit-cell volume evolution under pressure. Lines are the Birch-Mumaghan EoS fit. (Inset) Evolution of the $β$ angle of the monoclinic phase under pressure. The green region shows the pressure range of the phase coexistence. Measurements at T = 300 K.

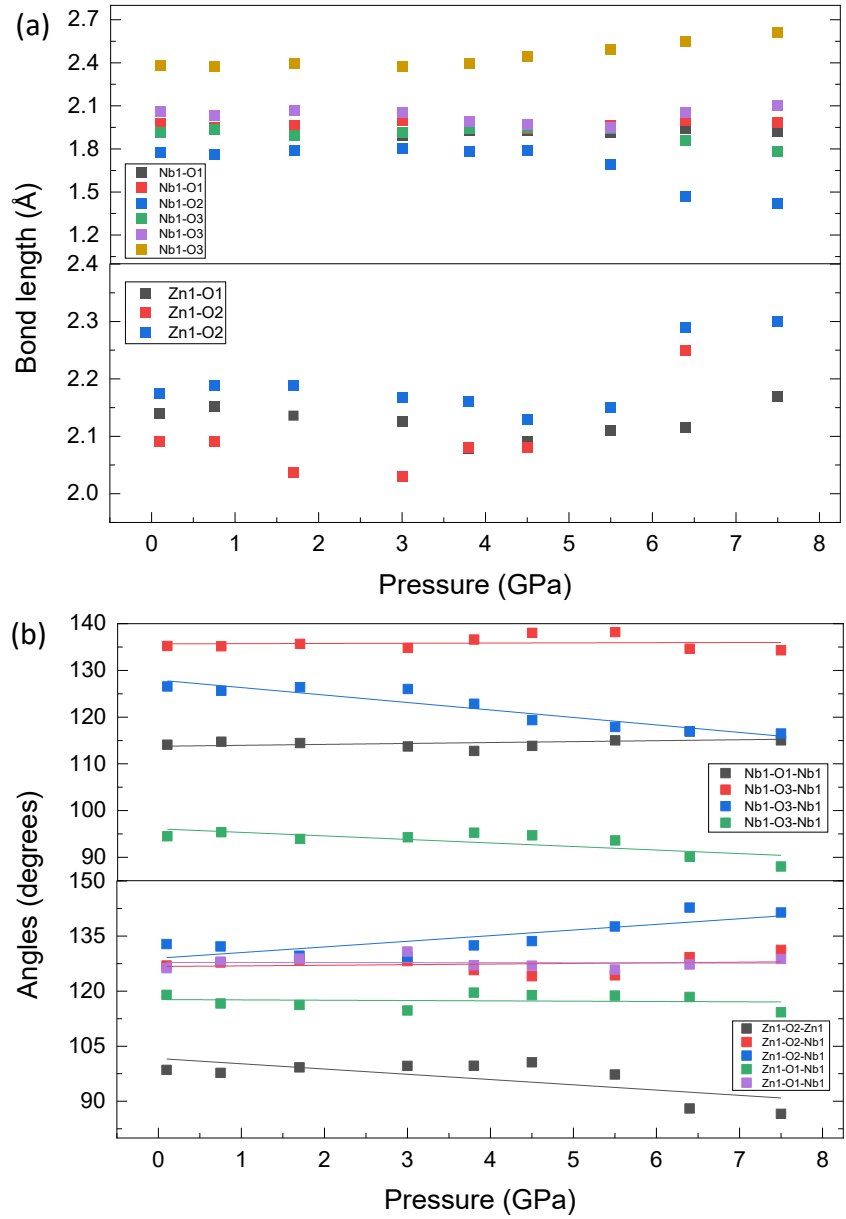

**Fig. 12.** (a) Bond length and (b) angles evolution for the two ZnO and NbO polyhedra under pressure at T = 300 K.

**Table 2:** Extended table from reference [16], for comparison of volume, bulk modulus and pressure derivative for various columbite structured compounds under quasi-hydrostatic pressure. The method and order of the equation of state are shown in parentheses.

| Compound | $V_0$ (Å3) | $B_0$ (GPa) | $B'_0$ | Reference |
|---|---|---|---|---|
| (Mn, Fe)Nb$_2$O$_6$,Fe/(Fe + Mn) (0.80) RAO no. 17 (exp. BM3) | 414.56 (2) | 149.0 (5) | 4.1 (2) | [10][9] |
| (Mn, Fe)Nb$_2$O$_6$,Fe/(Fe + Mn) (0.80) RAO no. 15 (exp. BM3) | 413.88 (4) | 153 (1) | 4.8 (3) | [10] |
| (Mn, Fe)Nb$_2$O$_6$,Fe/(Fe + Mn) 0.15 KRA no. 11 (exp. BM3) | 421.18 (3) | 146 (1) | 4.9 (4) | [10] |
| Mn(Ta,Ta)$_2$O$_6$ (exp. BM2) | 420 | 149(4) | 4 | [12] |
| MnNb$_2$O$_6$ (exp. BM3) | 424.2(3) | 154.4(3.5) | 4.1 | [11] |
| SnNb$_2$O$_6$ (DFT BM3) | 218.28 | 217 | 4 | [9] |
| NiNb$_2$O$_6$ (exp. BM3) | 399.98 | 178.7 (17) | 3.7 (5) | [16] |
| ZnNb$_2$O$_6$ (exp. BM3) | 408.5(2) | 165(7) | 2.7(15) | This work |
| ZnNb$_2$O$_6$ (exp. BM2) | 408.66(15) | 159.5(19) | 4 | This work |
| ZnNb$_2$O$_6$ (DFT BM3) | 422.39(5) | 142.9(9) | 4.01(13) | This work |
| ZnNb$_2$O$_6$ (DFT BM2) | 422.39(3) | 142.9(2) | 4 | This work |

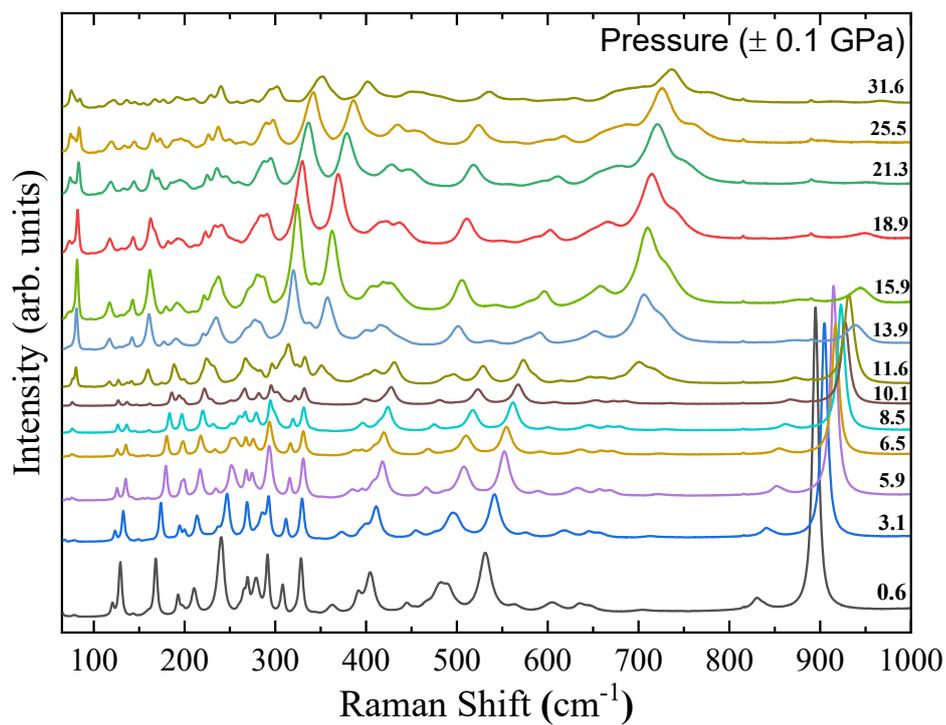

**Fig. 13.** Raman spectra of ZnNb$_2$O$_6$ at various pressures up to 31.6 GPa. Numbers on the right-hand side show the pressure in GPa. Measurements at T = 300 K.

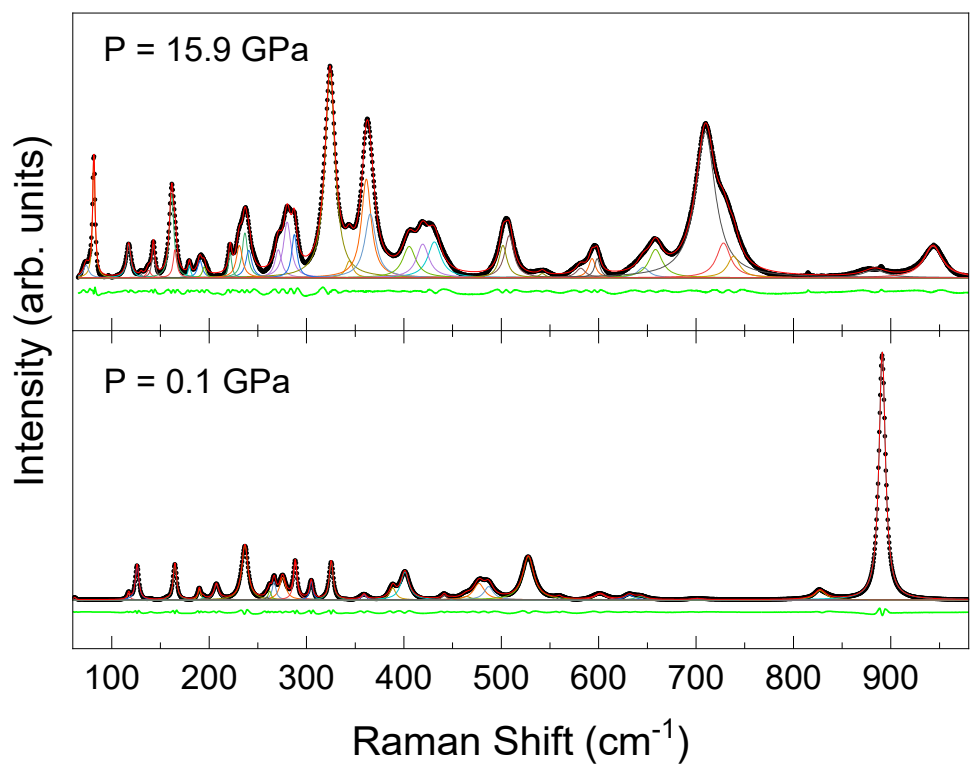

Fig. 14. Fitting of selected ZnNb$_2$O$_6$ Raman spectra of the low- and high-pressure phases. Measurements at T = 300 K.

**Table 3:** Raman active modes ($\omega_0$), pressure coefficients ($d\omega/dP$), Grüneisen parameters ($\gamma = (B_0/\omega_0)(d\omega/dP)$) and the corresponding symmetry of the orthorhombic ZnNb$_2$O$_6$ structure at ambient pressure. Experimental and theoretical values are reported. The experimental modes are tentatively assigned to the corresponding theoretical ones. $B_0 = 159.5(17)$ GPa from Table 1 was used to calculate $\gamma$. Present experimental frequencies are given with an accuracy of ±1 cm$^{-1}$ and pressure coefficients with an accuracy of ±0.5 cm$^{-1}$/GPa.

| No | Experimental | | | Theoretical | | | Mode |
|---|---|---|---|---|---|---|---|
| | $\omega$ (cm$^{-1}$) | $d\omega/dP$ (cm$^{-1}$/GPa) | $\gamma$ | $\omega$ (cm$^{-1}$) | $d\omega/dP$ (cm$^{-1}$/GPa) | $\gamma$ | |
| 1 | 62 | 1.3 | 3.44 | 63 | 0.9 | 2.16 | B$_{3g}$ |
| 2 | | | | 68 | 0.8 | 1.77 | A$_g$ |
| 3 | | | | 111 | 1.2 | 1.75 | B$_{2g}$ |
| 4 | | | | 113 | 1.6 | 2.22 | B$_{3g}$ |
| 5 | | | | 114 | 1.8 | 2.55 | B$_{1g}$ |
| 6 | 118 | 1.0 | 1.33 | 117 | 1.6 | 2.20 | A$_g$ |
| 7 | | | | 123 | 1.3 | 1.67 | B$_{2g}$ |
| 8 | 126 | 1.1 | 1.39 | 124 | 2.5 | 3.18 | B$_{1g}$ |
| 9 | 140 | 2.1 | 2.36 | 150 | 0.9 | 0.95 | B$_{3g}$ |
| 10 | 157 | 0.4 | 0.44 | 160 | 0.8 | 0.75 | B$_{1g}$ |
| 11 | | | | 163 | 1.3 | 1.29 | B$_{3g}$ |
| 12 | 165 | 2.1 | 2.04 | 167 | 2.0 | 1.90 | B$_{2g}$ |
| 13 | | | | 186 | 0.7 | 0.57 | A$_g$ |
| 14 | 190 | 0.7 | 0.59 | 188 | 0.6 | 0.50 | B$_{1g}$ |
| 15 | 194 | 0.3 | 0.25 | 196 | 0.4 | 0.30 | A$_g$ |
| 16 | 207 | 1.4 | 1.04 | 201 | 1.6 | 1.25 | B$_{3g}$ |
| 17 | | | | 218 | 0.7 | 0.51 | B$_{2g}$ |
| 18 | | | | 222 | 1.9 | 1.34 | A$_g$ |
| 19 | 237 | 1.7 | 1.16 | 229 | 1.6 | 1.12 | B$_{1g}$ |
| 20 | | | | 250 | 1.2 | 0.77 | B$_{1g}$ |
| 21 | | | | 253 | 0.9 | 0.57 | B$_{2g}$ |
| 22 | 259 | 5.9 | 3.64 | 253 | 1.1 | 0.68 | B$_{3g}$ |
| 23 | 262 | 0.4 | 0.25 | 265 | 0.2 | 0.10 | B$_{3g}$ |
| 24 | 267 | 1.3 | 0.79 | 268 | 2.2 | 1.28 | A$_g$ |
| 25 | | | | 268 | 2.2 | 1.29 | B$_{2g}$ |
| 26 | 276 | 2.3 | 1.31 | 279 | 1.0 | 0.58 | B$_{2g}$ |
| 27 | | | | 282 | 1.1 | 0.62 | A$_g$ |
| 28 | 288 | 1.0 | 0.54 | 294 | 1.4 | 0.78 | B$_{1g}$ |
| 29 | 305 | 1.6 | 0.83 | 313 | 0.7 | 0.35 | B$_{1g}$ |
| 30 | | | | 315 | 1.8 | 0.89 | A$_g$ |
| 31 | 325 | 0.6 | 0.28 | 332 | 1.4 | 0.68 | B$_{3g}$ |
| 32 | | | | 345 | 3.7 | 1.72 | B$_{3g}$ |
| 33 | 359 | 4.2 | 1.87 | 371 | 1.0 | 0.44 | B$_{1g}$ |
| 34 | | | | 375 | 2.9 | 1.23 | B$_{2g}$ |
| 35 | 389 | 3.1 | 1.28 | 382 | 2.6 | 1.10 | A$_g$ |
| 36 | | | | 398 | 4.4 | 1.78 | B$_{3g}$ |
| 37 | 401 | 2.6 | 1.05 | 413 | 3.9 | 1.51 | B$_{1g}$ |
| 38 | | | | 430 | 2.4 | 0.90 | B$_{3g}$ |
| 39 | | | | 430 | 4.1 | 1.53 | B$_{2g}$ |
| 40 | 441 | 3.9 | 1.42 | 448 | 4.8 | 1.72 | A$_g$ |
| 41 | 464 | 3.7 | 1.27 | 458 | 3.9 | 1.34 | B$_{2g}$ |
| 42 | 477 | 4.6 | 1.53 | 461 | 5.6 | 1.94 | B$_{1g}$ |
| 43 | 487 | 3.8 | 1.24 | 502 | 4.0 | 1.27 | A$_g$ |
| 44 | 528 | 3.9 | 1.19 | 536 | 4.8 | 1.44 | B$_{1g}$ |
| 45 | 560 | 4.9 | 1.38 | 568 | 5.2 | 1.47 | B$_{3g}$ |
| 46 | 602 | 5.1 | 1.36 | 607 | 3.9 | 1.03 | B$_{2g}$ |
| 47 | 632 | 4.0 | 1.02 | 620 | 4.0 | 1.03 | A$_g$ |
| 48 | 643 | 4.2 | 1.05 | 655 | 3.7 | 0.89 | B$_{1g}$ |
| 49 | 702 | 3.3 | 0.75 | 684 | 3.3 | 0.76 | B$_{3g}$ |
| 50 | | | | 789 | 3.8 | 0.77 | B$_{2g}$ |
| 51 | | | | 791 | 3.9 | 0.79 | B$_{1g}$ |
| 52 | 827 | 4.1 | 0.79 | 798 | 4.0 | 0.80 | B$_{3g}$ |
| 53 | | | | 850 | 3.5 | 0.66 | A$_g$ |
| 54 | 891 | 3.7 | 0.66 | 852 | 3.6 | 0.67 | B$_{2g}$ |

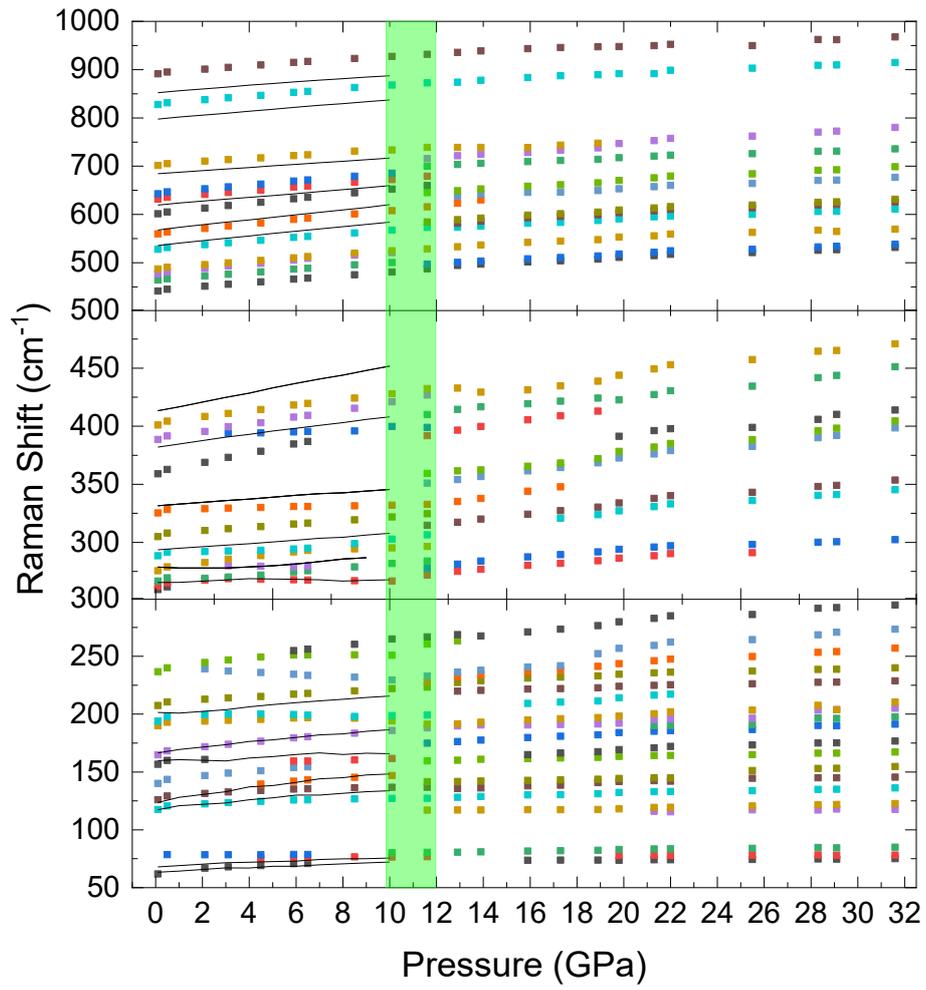

**Fig. 15.** Evolution of the Raman modes of ZnNb$_2$O$_6$ under pressure (T = 300 K). The green region shows the pressure region where the phase transition takes place. Solid lines represent selected DFT phonon calculations.

## Declaration of competing interest

The authors declare that they have no known competing financial interests or personal relationships that could have appeared to influence the work reported in this paper.

## Credit authorship contribution statement

**A. Tyagi:** Formal analysis, Methodology, Writing - review & editing.
**P. Botella:** Formal analysis, Writing - original draft, Writing - review & editing.
**A. B. Garg:** Conceptualization, Formal analysis, Writing - original draft, Writing - review & editing.
**J. Sánchez-Martín:** High-pressure High-temperature x-ray data collection.
**D. Díaz-Anichtchenko:** High-pressure High-temperature x-ray data collection.
**R. Turnbull:** High-pressure High-temperature x-ray data collection. Writing - review & editing.
**S. Anzellini:** High-pressure High-temperature x-ray data collection.
**C. Popescu:** High-pressure X-ray diffraction data collection
**D. Errandonea:** Conceptualization, Formal analysis, Writing - original draft, Writing - review & editing.

## Acknowledgements

D. E. thanks the financial support from the Spanish Ministerio de Ciencia e Innovación (https://doi.org/10.13039/501100011033) under Projects PID2019-106383GB-41, PID2022-138076NB-C41, and RED2022-134388-T. D. E. also thanks the financial support of Generalitat Valenciana through grants PROMETEO CIPROM/2021/075-GREENMAT and MFA/2022/007. This study forms part of the Advanced Materials program and is supported by MCIN with funding from the European Union Next Generation EU (PRTR-C17.I1) and by the Generalitat Valenciana. J. S.-M. acknowledges the Spanish Ministry of Science, Innovation and Universities for the PRE2020-092198 fellowship. R.T. acknowledges funding from the Generalitat Valenciana for Postdoctoral Fellowship No. CIAPOS/2021/20. S. A. thanks the Generalitat Valenciana for the CIDEGENT grant no. CIDEXG/2022/6. The authors thank ALBA for providing beamtime under experiment no. 2022085940

# Supplementary material

**Table S1**. DFT calculated Raman, IR, silent and acoustic modes for $ZnNb_2O_6$

| Raman | | Infrared | | Silent | | Acoustic | |
|---|---|---|---|---|---|---|---|
| ω (cm$^{-1}$) | Mode | ω (cm$^{-1}$) | Mode | ω (cm$^{-1}$) | Mode | ω (cm$^{-1}$) | Mode |
| 63.4 | B3g | 95.4 | B3u | 72.1 | Au | 0 | B2u |
| 68 | Ag | 110.6 | B2u | 143.3 | Au | 0 | B1u |
| 110.9 | B2g | 139.4 | B1u | 197.7 | Au | 0 | B3u |
| 113.1 | B3g | 151.1 | B2u | 216 | Au | | |
| 114.2 | B1g | 152.4 | B1u | 265.2 | u | | |
| 117.4 | Ag | 158 | B3u | 283.9 | Au | | |
| 122.8 | B2g | 172 | B3u | 318.8 | Au | | |
| 123.7 | B1g | 174.8 | B1u | 377.4 | Au | | |
| 150.1 | B3g | 194.1 | B1u | 416.3 | Au | | |
| 159.8 | B1g | 199.8 | B3u | 511 | Au | | |
| 162.8 | B3g | 203.5 | B2u | 538.3 | Au | | |
| 166.5 | B2g | 222.9 | B1u | 626.8 | Au | | |
| 185.5 | Ag | 228.1 | B3u | 764.6 | Au | | |
| 187.6 | B1g | 229.6 | B2u | | | | |
| 196.2 | Ag | 251.8 | B3u | | | | |
| 201.4 | B3g | 251.9 | B1u | | | | |
| 218.3 | B2g | 261 | B2u | | | | |
| 221.7 | Ag | 299.4 | B1u | | | | |
| 229.2 | B1g | 301 | B3u | | | | |
| 249.8 | B1g | 315.8 | B1u | | | | |
| 252.7 | B2g | 334 | B2u | | | | |
| 252.7 | B3g | 366.3 | B3u | | | | |
| 265.4 | B3g | 377.5 | B2u | | | | |
| 267.8 | Ag | 389.1 | B1u | | | | |
| 267.8 | B2g | 423.4 | B2u | | | | |
| 278.7 | B2g | 427 | B2u | | | | |
| 281.6 | Ag | 428.3 | B3u | | | | |
| 293.5 | B1g | 456.7 | B1u | | | | |
| 312.9 | B1g | 457.1 | B3u | | | | |
| 315.4 | Ag | 499.4 | B1u | | | | |
| 331.6 | B3g | 549.3 | B3u | | | | |
| 345.1 | B3g | 588.8 | B1u | | | | |
| 370.5 | B1g | 590.3 | B2u | | | | |
| 374.8 | B2g | 652.3 | B3u | | | | |
| 382 | Ag | 660.5 | B2u | | | | |
| 398.2 | B3g | 774.2 | B2u | | | | |
| 413.3 | B1g | 797.8 | B3u | | | | |
| 429.5 | B3g | 821 | B1u | | | | |
| 430.4 | B2g | | | | | | |
| 447.8 | Ag | | | | | | |
| 458.4 | B2g | | | | | | |

| | |
|---|---|
| 461.4 | B1g |
| 501.5 | Ag |
| 535.9 | B1g |
| 567.9 | B3g |
| 606.8 | B2g |
| 619.5 | Ag |
| 654.6 | B1g |
| 684.4 | B3g |
| 789.4 | B2g |
| 791 | B1g |
| 797.6 | B3g |
| 850.1 | Ag |
| 852.1 | B2g |

Equation of State Study

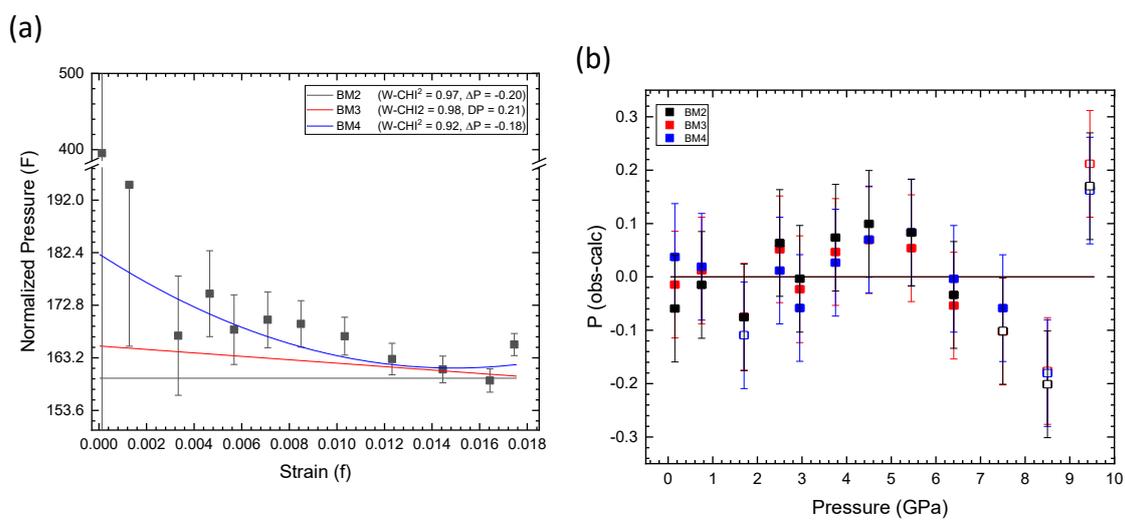

**Fig. S1.** (a) F-f plot for the three BM equation fits. (b) Pobs-Pcalc after fitting the data with the three BM equations. Empty squares data not touching the reference line.

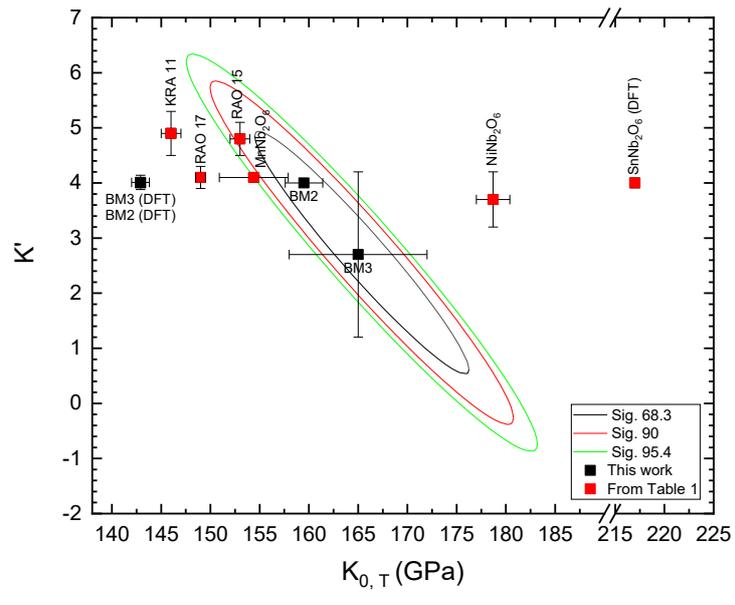

**Fig. S2.** Confidence ellipses from the present results and data from Table 1.